\documentclass[a4paper,twocolumn,english]{revtex4}
\usepackage[T1]{fontenc}
\usepackage[latin9]{inputenc}
\usepackage{amsthm}
\usepackage{amsmath}
\usepackage{graphicx}
\usepackage{amssymb}
\usepackage{esint}

\makeatletter
\@ifundefined{textcolor}{}
{%
 \definecolor{BLACK}{gray}{0}
 \definecolor{WHITE}{gray}{1}
 \definecolor{RED}{rgb}{1,0,0}
 \definecolor{GREEN}{rgb}{0,1,0}
 \definecolor{BLUE}{rgb}{0,0,1}
 \definecolor{CYAN}{cmyk}{1,0,0,0}
 \definecolor{MAGENTA}{cmyk}{0,1,0,0}
 \definecolor{YELLOW}{cmyk}{0,0,1,0}
 }

\makeatother

\usepackage{babel}

\begin{document}

\title{Phospholipid membranes repulsion at $nm$-distances explained within
a continuous water model.}

\author{P. O. Fedichev }

\affiliation{$^{1)}$Quantum Pharmaceuticals Ltd, Ul. Kosmonavta Volkova 6A-606,
125171, Moscow, Russian Federation}

\email{peter.fedichev@q-pharm.com}

\homepage{http://q-pharm.com}

\author{L. I. Menshikov}

\affiliation{$^{2)}$RRC Kurchatov Institute, Kurchatov Square 1, 123182, Moscow,
Russian Federation}
\begin{abstract}
We apply a recently developed phenomenological theory of polar liquids
to calculate the repulsive pressure between two hydrophilic membranes
at $nm$-distances. We find that the repulsion does show up in the
model and the solution to the problem fits the published experimental
data well qualitatively and even quantitatively. Moreover, we find
that the repulsion is practically independent of the temperature,
and thus put some extra weight in favor of the so called hydration
over entropic hypothesis for the membranes interactions explanation.
The calculation is a good {}``proof of concept'' example a sufficiently
sophisticated continuous water model application to a non-trivial
interaction on $nm$-size objects in water arising from long-range
correlations between the water molecules. 
\end{abstract}
\maketitle
The solvent induced forces play an important role in Nature, key nano-
and bio-technological applications, drive various self-assembly phenomena
in cells and membranes \cite{chandler2005interfaces}, vesicle-membrane,
and -synapse fusion \cite{ahkong1975mechanisms,pagano1974interaction},
protein folding \cite{Lum}. Understanding of these phenomena is crucial
in a number of practically important applications such as drug design,
drug transport optimization, and the design of nano-particles drug
transport and delivery systems \cite{ranade2003drug}. 

One of the most conceptually simple while still an important and an
easily experimentally observable example of such interactions is provided
by the short range repulsion between phospholipid membranes, originally
discussed in \cite{leneveu1976measurement,sackmann1995structure}.
Since the original discovery there has been a lot of efforts to understand
the nature of the repulsive forces, as described in \cite{sackmann1995structure}
and the refs. therein. Most of the time the interactions are analyzed
within some kind of a two-body approximation in such a way that the
total water mediated pressure between the two parallel lipid membranes
is represented as a the sum of the two components: the direct pressure,
arising from the direct interactions between the opposing membranes,
and the hydration component, associated with the interaction of the
membranes with the intervening water molecules \cite{pertsin2007temperature}.

Since the membranes are hydrophilic, the exclusion of the water leads
to the energy loss, and hence on the molecular level the hydration
pressure can be associated with the orientation polarization of water
near the membrane surfaces \cite{rand1989hydration,leikin1993hydration,marcelja1976repulsion}.
An alternative approach summarized in \cite{israelachvili1996role}
suggests the dominant role the water molecules ordering next to the
surfaces and thus emphasizes the water entropy contribution to the
free energy of the system. None of the effects are easy to grasp within
any kind of a simple continuous water model, which are normally designed
to reproduce the effects of continuous electrostatics in polar solutes,
e.g. \cite{baker2001ena,schafer2000cit,tjong2007gbr6}. The reason
is that due to long-range electrostatic interactions between the molecular
dipoles the correlations in water are collective and survive at very
long distances up to $1nm$. Therefore the effects of the ordering
may lead to cluster formations and phase transitions phenomena \cite{angell1973anomalous,hodge1978relative,speedy1976isothermal,ter1981thermodynamic,oleinikova2005formation,oleinikova2005percolation,fedichev2008fep},
appearance of strong and sufficiently long range interactions of non-electrostatics
nature \cite{fedichev2006long,Nechaev,Lum}. 

The alternative to the continuous solvation models is Molecular Dynamics
(MD) \cite{rapaport2004amd} of the body of interest immersed in a
tank of water molecules in a realistic force field or even within
quantum mechanical settings. Though such an approach may in principle
provide ultimately accurate predictions, the calculations are time
consuming and pose a number of challenges stemming, e.g. from long
relaxation times of water clusters. One possible way to bridge the
{}``simulation gap'' is to develop advanced of continuous solvation
models, such as \cite{fedichev2006long,fedichev2008fep,gong2008influence,gong2009langevin}
and test its limits to make sure the models include realistic and
important interactions. In what follows we take the polar liquid phenomenology
and calculate the repulsive pressure between two hydrophilic membranes
at $nm$-distances. We find that the repulsion does show up in the
continuous model and that the solution to the problem fits the available
the published experimental data well both qualitatively and quantitatively.
The pressure turns out to be practically independent of the temperature,
which strongly supports the hydration nature of the water-assisted
interactions between the membranes. The interaction appears to be
generic, should arise between any hydrophilic bodies and vanishes
as soon as the objects in question are separated farther then $R_{D}\sim1nm$,
the characteristic polarized water domain size identified in the model. 

The continuous polar liquid model introduced in \cite{fedichev2006long}
is capable of describing both the short- and the long-range features
of a polar liquid in a single theoretical framework. Originally it
was applied to calculate water-assisted interactions of macroscopic
bodies with hydrophobic interfaces of various shapes and charges.
It was shown that the competition between the short range (hydrogen
bonding) and the long-range dipole-dipole interactions of the solvent
molecules leads to appearance of strong, long range and orientationally
dependent interactions between the objects, which, in principle, can
be responsible for various self-assembly processes in biological systems.
Within the suggested model a polar liquid itself is characterized
by a complicated fluctuating thermal state, ordered at sufficiently
short scales within a single domain, and completely disordered at
larger distances. This physical picture has far reaching consequences,
especially at solvent-solute surfaces, where the ferroelectric film
of solvent molecules may be formed \cite{men2009spontaneous}. In
accordance with the MD simulations \cite{oleinikova2005formation,oleinikova2005percolation}
the vector model predicts the BKT-like topological phase transitions
at solvent-solute interfaces \cite{men2009nature,fedichev2009bkt}.
Most of our earlier research was confined to interactions of hydrophobic
bodies. In this Paper we try the model and provide the solution to
the phospholipid membranes repulsion problem, which is a hydrophilic
bodies interaction example.

The model description of a polar liquid, e.g. water, proceeds as follows.
Each of the molecules within the liquid ($j=1,2,\ldots,N$) is assumed
have a vector property: the static electric dipole moment $\mathbf{d}_{j}=d_{0}\mathbf{S}_{j}$,
where $d_{0}$ is its magnitude, and $\mathbf{S}_{j}$ is the unit
vector in expressing the orientation of the molecule. Having this
in mind it is possible to develop a vector field theory in which the
liquid is described by a local mean value of the molecular polarization
vector $\mathbf{s}(\mathbf{r})=\langle\mathbf{d}\rangle/d_{0}$, where
$\langle\mathbf{d}\rangle\equiv\langle\mathbf{d}_{j}\rangle$ stands
for the statistical average of molecular dipole momenta $\mathbf{d}$
over a small but sufficiently large volume of the liquid containing
macroscopic number of molecules. Accordingly, the model expression
for the free energy functional is given by: \[
G[\mathbf{s\left(\mathbf{r}\right)}]=P_{0}^{2}\int dV\frac{C}{2}\sum_{\alpha,\beta=x,y,z}\frac{\partial s_{\alpha}}{\partial x_{\beta}}\frac{\partial s_{\alpha}}{\partial x_{\beta}}+\]
\begin{equation}
+P_{0}^{2}\int dVV(\mathbf{s}^{2})+{\displaystyle \int dV\frac{1}{8\pi}\mathbf{E}_{P}^{2}}-\int dV\mathbf{P\left(\mathbf{r}\right)E_{\mathbf{e}}\left(\mathbf{r}\right)}.\label{eq: Free energy simple}\end{equation}
The polarization vector of the liquid, $\mathbf{P}=P_{0}\mathbf{s}$
where $P_{0}=n_{0}d_{0}$, $n_{0}$ is the molecular density, and
$\mathbf{E}_{e}=-\nabla\varphi_{e}$ is the external electric field
in the absence of the liquid. Similarly, $\mathbf{E}_{P}=-\nabla\varphi_{p}$
is the polarization electric field, produced by the polarization charges
within the liquid characterized by the polarization charge density
$\rho_{P}=-{\rm div}\mathbf{P}$. The polarization potential $\varphi_{P}$
should be found from the Poisson equation $\triangle\varphi_{p}=-4\pi\rho_{p}$. 

The Oseen's like term in Eq.(\ref{eq: Free energy simple}) provides
a model description of the hydrogen bonds network deformation energy
in the long wavelength limit. The value of the phenomenological parameter
$C\approx10^{-15}cm^{2}$ , as well as the specific form and characteristic
values of the function $V\left(s^{2}\right)$ describing the elastic
energy of water polarization should be extracted from the experimental
properties of the liquid as described in in \cite{fedichev2006long}.
Typically the liquid polarization is small, $s\ll1$, and {}``the
equation of state'' function takes the usual Ginzburg-Landau form
\begin{equation}
V(\mathbf{s}^{2})\approx\frac{A}{2}s^{2}+Bs^{4},\label{eq:V at small s}\end{equation}
where $A=4\pi/\left(\varepsilon-1\right)\approx0.16$, $\varepsilon\approx80$
is the (large) dielectric constant, and $B\sim1$. The parameter $A$
characterizes the long-range interactions of the molecular dipoles,
depends strongly on the temperature: $A\varpropto\left(T-T_{C}\right)/T_{C}$
\cite{fedichev2008fep}, and vanishes at $T=T_{C}\approx228K$ roughly
at the $\lambda-$transition point in supercooled water \cite{angell1973anomalous,hodge1978relative,speedy1976isothermal,ter1981thermodynamic}.
On the contrary, the parameter $B$ depends on the short-range physics
and thus is practically independent of the temperature. The last two
terms in Eq. (\ref{eq: Free energy simple}) describe the long-range
dipole-dipole interaction of molecules, and interaction energy of
the liquid in the external electric field, produced by charges that
reside inside or outside the liquid. 

\begin{figure}
\includegraphics[width=0.9\columnwidth]{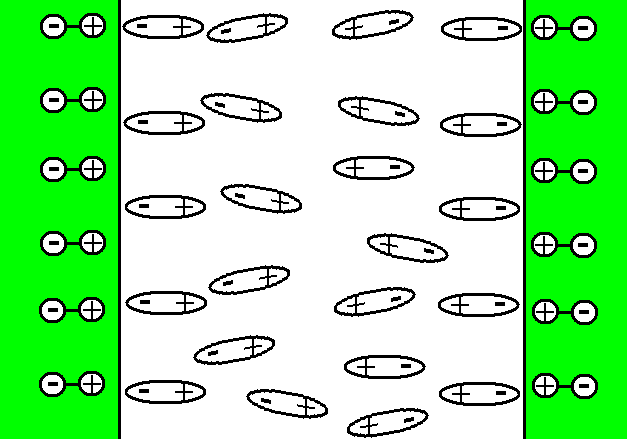}

\caption{Typical polarization of the water molecules between lipid membranes
shown in green. Both the charges of the lipid polar groups and the
polarization of the water molecules are schematically represented
by the corresponding charge symbols. \label{fig:Polarization-of-water}}

\end{figure}

The cell membrane consists primarily of a thin layer of amphipathic
phospholipids which spontaneously arrange so that the hydrophobic
``tail'' regions are shielded from the surrounding
polar fluid, causing the more hydrophilic ``head''
regions to associate with the cytosolic and extracellular faces of
the resulting bilayer, as shown on Fig. (\ref{fig:Polarization-of-water}).
Typically the width of the lipid bilayer is up to $4nm$. Consider
two membranes of the cross section area $S$ separated by the water
filled layer of width $h$ from each other. Typically the membrane
surfaces are comprised by molecular groups with no net charge, though
with a considerable dipole moment. The lipid {}``heads'' are hydrophilic,
therefore water molecular dipoles arrange themselves in the direction
of the lipid wall: $s(\pm h/2)=\pm s_{0}$, $s_{0}\sim1$ (of course,
$0<s_{0}<1$: the value of $s_{0}$ depends on the dipole moments
of lipid polar groups). 

Since there is no external electric field in our system, $\mathbf{E}_{P}=-4\pi\mathbf{P}=-4\pi P_{0}\mathbf{s}$
and the expression for the free energy of the liquid reads\begin{equation}
G=SP_{0}^{2}\int dz\left(\frac{C}{2}\left(\frac{ds}{dz}\right)^{2}+U(s)\right),\label{eq:total_energy}\end{equation}
where $s=\mathbf{s}_{z}$ is the only non-vanishing component of the
average water molecules polarization and $U(s)=V(s^{2})+2\pi s^{2}$. 

Variational minimization of the free energy (\ref{eq: Free energy simple})
with respect to the variations of the function $s\left(\mathbf{r}\right)$,
$\delta G=0$, gives the equation\begin{equation}
-\frac{C}{2}\frac{d^{2}s}{dz^{2}}+\left(V^{\prime}(s^{2})+2\pi\right)s=0\label{eq: Equation of motion}\end{equation}
similar to the equation of motion of a particle of the mass $C$ at
the position $s$ moving in one-dimensional quasi-potential $U(s)$.
Accordingly, the derivative $ds/dz$ plays the role of the particle
velocity velocity and the variable $z$ serves as the time. The solution
is a well known 1d soliton\[
\frac{ds}{dz}=\pm\sqrt{\frac{2}{C}\left(\epsilon+V(s^{2})+2\pi s^{2}\right)},\]
where the exact value of the {}``energy'' is a constant to be found
by matching the boundary conditions at the membrane surfaces $s(\pm h/2)=s_{0}$:\begin{equation}
\frac{h}{\sqrt{2C}}=\int_{0}^{s_{0}}\frac{ds}{\sqrt{\epsilon+U(s^{2})}}.\label{eq:boundary_integral}\end{equation}
The function$\epsilon(h)$ simplifies in the large $h$ limit: the
main contribution to the integral comes the small $s$ region where
the integral diverges logarithmically, the function $V(s^{2})$ can
be neglected altogether, $U(s^{2})\approx2\pi s^{2}$, and therefore\begin{equation}
\epsilon\approx2\pi s_{0}^{2}\exp(-h/\lambda),\label{eq:epsilonofh}\end{equation}
where $\lambda=\sqrt{C/4\pi}\ll h$ is the characteristic size of
the soliton, $\lambda\sim R_{D}\approx0.15\div0.25nm$, the characteristic
water orientation domain size, first introduced in \cite{fedichev2006long,fedichev2008fep}. 

Eq. (\ref{eq:total_energy}) can be used to transform the energy of
the liquid layer to \[
\frac{G}{SP_{0}^{2}}=2C\int_{0}^{h/2}dz\left(\frac{ds}{dz}\right)^{2}-\varepsilon h=\]
\[
=\sqrt{2C}\int_{0}^{s_{0}}ds\sqrt{\epsilon+U(s^{2})}-\varepsilon h.\]
According to the standard definition the pressure, $P=-S^{-1}\partial G/\partial h$,
can be expressed as:\[
P=-\sqrt{2C}P_{0}^{2}\frac{d\epsilon}{dh}\int_{0}^{s_{0}}\frac{ds}{\sqrt{\epsilon+U(s^{2})}}+P_{0}^{2}(\varepsilon+h\frac{d\epsilon}{dh}).\]
The integral in the r.h.s. can be evaluated using Eq. (\ref{eq:boundary_integral})
and the asymptotic expression (\ref{eq:epsilonofh}) for $\epsilon(h)$,
so that the pressure is given by\[
P=P_{1}\exp(-h/\lambda),\]
where the prefactor $P_{1}=2\pi P_{0}^{2}s_{0}^{2}$. This is exactly
the dependence observed in experiments \cite{leneveu1976measurement}.
The pressure is indeed positive, the membranes do repel each other,
and the forces vanish exponentially quickly as soon as the membranes
depart further apart than $h\sim\lambda$. 

The physical reason of repulsion is the formation of polarization
charges in the middle of the water layer due to inhomogeneous water
molecules alignment. The polarization charges are all of the same
sign (positive for the example presented on Fig.\ref{fig:Polarization-of-water})
and thus generate the electrostatic repulsion, which together with
the hydrogen bonds network deformation energy is ultimately responsible
for the repulsion of the membranes. For the maximum possible surface
polarization, $s_{0}=1$, which corresponds to the large dipole moments
of the lipid polar groups limit, we can estimate the maximum value
of the prefactor: $\left(P_{1}\right)_{max}\approx3\cdot10^{10}$
$dyne/cm^{2}$. The value of the pressure obtained in this way agrees
well with the results of the measurements reported for different types
of the membranes: $5\cdot10^{9}<\left(P_{1}\right)_{exp}<5\cdot10^{10}$
$dyne/cm^{2}$ \cite{leneveu1976measurement}.

Few concluding remarks should be added here. First, the repulsion
pressure depends essentially only on $s_{0}=\left\langle S_{z}\right\rangle _{\Gamma}$,
$z$-component of unit vector $\mathbf{S}$ directed along the dipole
moment of water molecule on the membrane interface $\Gamma$. This
quantity decreases slightly as the temperature increases in agreement
with the molecular dynamics studies of \cite{pertsin2007temperature}.
This observation lets us put some more weight in favor of the polarization
\cite{rand1989hydration,leikin1993hydration,marcelja1976repulsion}
over the so called {}``entropy'' hypothesis \cite{israelachvili1996role}
of the repulsion pressure. Indeed, the polarization pressure $P$
decreases slowly as the temperature rises, whereas any entropy-related
effect should lead to the sharp increase of $P$, as explained in
\cite{pertsin2007temperature}. 

Second, the model defined by Eq. (\ref{eq: Free energy simple}) is
very similar to the non-linear screening model introduced in \cite{gong2008influence,gong2009langevin}
and was originally applied for the calculation of the electrostatic
forces in water. The non-linear screening model does not contain the
Oseen term responsible for the short-range ordering of the water molecules.
The scale $\lambda$ is the characteristic size of the water molecules
domain (cluster) depends on $C$ and thus can only appear in the complete
model (\ref{eq: Free energy simple}). This makes our model apparently
the minimal continuous model capable of predicting repulsion of hydrophilic
membranes. We note that the separate problem interactions of hydrophobic
objects has been also extensively studied within another class of
two-scale continuous water models \cite{Chandler,Wolde,Nechaev}.
We leave the research on possible convergence of the approaches and
the relation between the scales of the models for a future publication.

The work was supported by Quantum Pharmaceuticals. Phenomenological
water models are used in Quantum software to compute solvation energies,
protein-drug interactions and drug transport properties in aqueous
environments.

\end{document}